\documentclass[12pt]{iopart}
\usepackage{graphicx,color,iopams}
\newcommand{\figsch}{%
\begin{figure}[htbp]
   \includegraphics[clip]{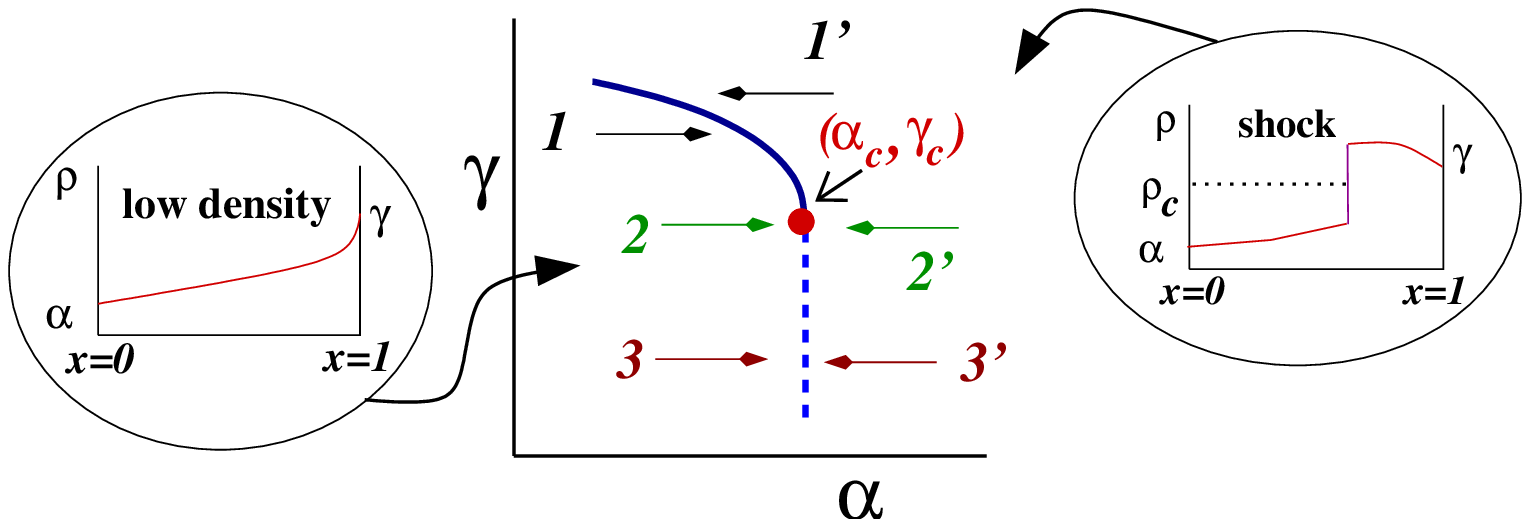}
    \caption{    Bulk phase
      boundary in the $\alpha-\gamma$ plane, separating a
      low-density phase and a phase with a shock (insets show the
      density variation $\rho$ vs $x$).  The critical point is at
      $(\alpha_c,\gamma_c)$.  The shock has a nonzero height on the
      solid line of the phase boundary, $\gamma>\gamma_c$ but the
      height vanishes on the dashed part, $\gamma<\gamma_c$.  Arrows
      indicate various paths used in the text.}
\label{fig:sch}
\end{figure}
}

\begin{document}
\letter{Nonequilibrium Criticality at Shock Formation in steady states}
\author{Sutapa Mukherji}
\address{Department of Physics, Indian Institute of
  Technology, Kanpur 208 016, India}
\eads{\mailto{sutapam@iitk.ac.in}}
\author{Somendra  M. Bhattacharjee} 
\address{ Institute of Physics, Bhubaneswar -751005, India}
\eads{\mailto{somen@iopb.res.in}}
\date{\today}
\begin{abstract}
  The steady-state shock formation in processes like nonconserving
  asymmetric simple exclusion processes in varied situations is shown
  to have a precursor of a critical deconfinement transition on the
  low-density side.  
 The diverging length scales and the quantitative
  description of the transition are obtained from a few general
  properties of the dynamics without relying on specific details.
\end{abstract}
 
\maketitle
 
Asymmetric simple exclusion processes (ASEP) involve particles hopping
in a preferred direction under hard-core repulsion that forbids double
occupancy on a site \cite{schuetzrev}.  This model with periodic
\cite{ligett,goutam} and open boundary conditions \cite{derrida1} as
well as its variants involving different update schemes \cite{ evans1}
have been extensively studied in order to gain general understanding
of far-from-equilibrium processes.  In addition to this, ASEP has
direct resemblance with the transport processes within the cell
provided the dynamics of ASEP is modified by allowing attachment and
detachment of particles from or to the environment respectively.  The
motor proteins which participate in cellular transportation by moving
on linear tracks laid by long bio-molecules play the role of particles
in ASEP \cite{alberts,frey,klumpp,levine}.

In the presence of open boundaries, one needs to think of two particle
reservoirs attached to the boundaries which either inject or withdraw
particles to or from the boundaries with certain specified rates.  An
additional reservoir is needed for the desorption/adsorption kinetics
(Langmuir Kinetics) of particles on the lattice.  The biased hopping
of the particles, injected at one end by the reservoir, causes a
finite current in the system even in the steady state.  It is
intuitively understandable that this particle current would help in
the propagation of the boundary information to the bulk of the system.
Thus unlike equilibrium systems, here boundaries play a crucial role
in the steady-state dynamics and can give rise to several new features
such as boundary-driven phase transition or production of shocks in
the density profile\cite{krug,kolo}.

The boundary-related event, that concerns us here is the appearance of
localized shocks in the density profile.  Various aspects of shocks,
which are discontinuities in the particle density profile over a
microscopic distance in the bulk, have been extensively studied in the
past.  If $\alpha$ and $\gamma$ are the densities maintained by the
reservoirs at the two ends, then in the $(\alpha,\gamma)$ plane there
are lines $\alpha=\alpha_s(\gamma)$ demarcating the possible phases.
See Fig. \ref{fig:sch}.  Such bulk phase diagrams are now known for
many cases and in fact  mean-field descriptions seem to give a good
description of the bulk phase diagrams, especially for the shock
formation\cite{krug,frey,evans,popkov}.  Here we show the existence of
a novel deconfinement transition of a layer near an open end as the
phase boundary is approached from the low-density side, reminiscent of
the equilibrium wetting transition\cite{schick}.  This layer with a
non-bulk density profile remains attached to the end point, but, after
deconfinement, admits the bulk density variation though with a shock.
Let us call this special layer a shockening layer and the transition a
``{\it shockening}'' transition.

\figsch

The shockening transition on the low density side is a precursor to
the bulk phase transition, and shows power-law behaviors.  This
criticality is characterized by two length scales $\xi$ and $w$, where
$\xi\to\infty$ leads to deconfinement while $w$ gives the length scale
for the crossover of the surface density profile to the bulk.  Though
$w$ in general remains finite, there is a possibility of $w\to\infty$
which would signal a criticality at $(\alpha_c,\gamma_c)$ on the bulk
phase transition line.  These divergences are described by the
exponents $\zeta_{-}$ and $\zeta_{c}$ defined by
%\begin{subequations}
\begin{eqnarray}
  \label{eq:4} 
  \xi&\sim& | \Delta\alpha|^{-\zeta_{-}}, \quad {\rm for}
     \quad \Delta\alpha\equiv \alpha -\alpha_s(\gamma) \rightarrow 0-,\\
  w&\sim& | \Delta\alpha|^{-\zeta_{c}}, \quad {\rm for} \quad
     \Delta\alpha \equiv \alpha-\alpha_c \rightarrow 0-,
\label{eq:4w}
\end{eqnarray}
%\end{subequations}
where $\Delta\alpha$ measures the deviation from the phase boundary
for a fixed $\gamma$. In equation (\ref{eq:4}) it is along a path like path
1 in Fig. \ref{fig:sch}  with $\gamma\neq\gamma_c$ while for equation
(\ref{eq:4w}) it is for $\gamma=\gamma_c$ (path 2 in Fig.
\ref{fig:sch}).  By analysing a general equation for the steady state,
we show that these two exponents $\zeta_{-}$ and $\zeta_{c}$ (and, in
fact, several other bulk exponents defined below) are {\it universal}
as they are determined by only a few general properties of the
dynamics and not on details.  

A very well-studied example is the case of nonconserving ASEP of one
species on a lattice of $N$ sites.  The particles can jump to the
neighboring forward site if it is empty. Apart from that, the bulk of
the system is attached to a particle reservoir such that a particle
can attach to (detach from) the chain with a rate
$\omega_a$($\omega_d$).  The dynamics at the left boundary is that of
the injection of particles with a rate $\alpha$ while at the right it
is withdrawal at the rate $1-\gamma$.  Since the mean-field dynamics
through average density like variables gives a good 
description of the bulk phase diagram,   we  adopt the same approach to
study the shockening transition.  This is expected to capture the
overall features of the transition though fluctuations may affect the
exponents.  The role of fluctuations will be studied elsewhere.

The mean-field equation describing the evolution of the particle
density at a site $i$, is expressed in terms of the density
$n_i=<\tau_i>$, where $\tau_i$ is the occupation number of site $i$
and $\langle ...\rangle$ denotes statistical average.  The dynamics is
given by
\begin{equation}
  \label{eq:12}
 \frac{dn_i}{dt}=
n_{i-1}(1-n_i)-n_i(1-n_{i+1})+\omega_a (1-n_i)- \omega_d n_i
\end{equation}
This dynamics can obviously be extended to incorporate other effects
as well.  In the large-$N$ limit, a continuum mean field approach is
based on the density variable $\rho(x)$ related to $\tau$ as $\langle
\tau_{i\pm 1}\rangle= \rho(x)\pm \frac{1}{N} \frac{\partial
  \rho}{\partial x}+ \frac{1}{2N^2} \frac{\partial^2\rho}{\partial
  x^2}....$ treating $1/N$ as the lattice spacing with $x$ in the
range $[0,1]$ .  The various forms of single species dynamics studied
so far can be written in a general form (for the steady state)
\begin{equation}
  \label{eq:25}
  \epsilon\,\frac{d }{d x} f_2(\rho) \frac{d \rho}{d x} +
  f_1(\rho) \frac{d \rho}{d x} + \Omega f_0(\rho) =0,
\end{equation}
with $f_{i}(\rho)$, $i=0,1,2$, specifying the dynamics of the system
and $\Omega=\omega_{d} N$.  The
steady state density profile satisfies the two boundary conditions
$\rho(x=0)=\alpha$ and $\rho(x=1)=\gamma$ at the two ends. Here
$\epsilon\equiv (2N)^{-1}$ is the small parameter.  
For example, for the case of noninteracting particles in equation
 (\ref{eq:12}) the
$f$-functions of equation (\ref{eq:25}) are 
\begin{eqnarray}
  \label{eq:2}
f_2(\rho)=1,\  
f_1(\rho)=2\rho -1,\ 
f_0(\rho)=K  (1-\rho)- \rho, (K=\omega_a/\omega_{d}).
\end{eqnarray}
The complexity of the functions $f_i$'s increases with interaction and
other details but  explicit knowledge of these functions is not
essential for the analysis reported here.  

The dynamics of equation (\ref{eq:25}) admits two special densities.  (i)
The Langmuir density corresponding to $\rho=\rho_L$ at which
$f_0(\rho_L)=0$; $\rho(x)=\rho_L$ is a particular solution of equation
(\ref{eq:25}). It represents the steady-state bulk density if
adsorption/desorption were the sole dynamics in the problem.  (ii) A
density $\rho=\rho_c$ at which $f_1(\rho)=0$. In case of a simple
zero, for small deviation $\rho=\rho_c+\delta\rho$, the hopping rules
show a special symmetry of invariance of the first derivative term
under $\delta\rho \to-\delta\rho$.  This is the particle-hole symmetry
of equation  (\ref{eq:2}).  The bulk dynamics may not respect this symmetry
in the general case of $\rho_c\neq \rho_L$.  The special symmetry
occurs for $\rho_c=\rho_L$ as e.g.  for equation (\ref{eq:2}) when $K=1$.
We take $f_2(\rho)\neq 0$ for $0\leq \rho\leq1$.  The cases of
non-simple zeros or zeros of $f_2(\rho)$ are to be discussed
elsewhere.

To study the shockening transition, we generate a uniform
approximation of the solution of equation  (\ref{eq:25}) 
via a leading-order boundary layer 
analysis\cite{cole}.  In general both the
boundary conditions cannot be satisfied if the second derivative term
($\epsilon\rightarrow 0$) is ignored.  As a result there appears a
boundary layer at one end or a shock somewhere in the interior (or
both).  Within this special region, the second derivative term is
needed.  By neglecting appropriate terms from the original equation,
one obtains two different solutions, to be called the outer and the
inner solutions such that the outer solution is valid over almost the
entire system and the inner solution is valid only in the region where
the boundary layer or the shock appears. The two solutions join
smoothly.  As a general rule, if the inner solution attains a
saturation then the boundary condition might not be satisfied by the
inner solution.  This is the criterion for shock formation.  In this
situation, the two boundary conditions are satisfied by two different
outer solutions connected by the inner solution in the interior
forming a shock layer.

Without loss of generality we choose the boundary or shock layer to be
at or near $x=1$.  For equation (\ref{eq:25}) the outer and inner solutions
come from
\begin{equation}
\label{eq:14}
\frac{d\rho_{\rm out}}{dx}= -\Omega \,
\frac{f_0(\rho_{\rm out})}{f_1(\rho_{\rm out})},\quad
{{\rm and}} \quad
\frac{d\rho_{\rm in}}{d\tilde x}=
\frac{F(\rho_{\rm in})}{f_2(\rho_{\rm in})}, 
\end{equation}
where $\tilde x=(x-x_d)/\epsilon$ is the inner variable, $x_d$ giving
the location of the layer or the shock, and
\begin{equation}
F(\rho)\equiv {\hat{f_1}}(\rho_{\rm o}) - {\hat{f_1}}(\rho),
\ {\rm with}\ \frac{d{\hat{f_1}}(\rho)}{d\rho}=f_1(\rho).
\label{eq:14a}
\end{equation}
The matching condition $\rho_{\rm
  in}(\tilde{x}\rightarrow-\infty)=\rho_{\rm o}\equiv \rho_{\rm
  out}(1)$, for smooth joining has been incorporated in equation
(\ref{eq:14a}).

Given that $\rho(0)=\alpha$ the  relevant  inner and outer solutions
are 
\begin{equation}
\label{eq:16}
\rho_{\rm in}(\tilde{x})=\rho_{\rm o}\, S_{\rm in}(\tilde x/w +\xi),
\ {\rm and}\ \Omega x= g(\rho_{\rm out})-g(\alpha) \quad ({\rm left \ solution})\\
\end{equation}
%and the inner solution can  be written as
%\begin{equation}
%\rho_{\rm in}(\tilde{x})=\rho_{\rm o}\, S_{\rm in}(\tilde x/w +\xi),
%\label{eq:sin}
%\end{equation}
with $S_{\rm in}(\tilde x)\to 1$ as $\tilde x\to -\infty$. 
The functional forms of $g(\rho)$  and $S_{\rm in}(\tilde x)$ depend
on the details of  $f$-functions.
By choosing the shift $x_d=1$, $\xi$ is determined by
\begin{equation}
  \label{eq:3}
  \rho_{\rm o}\, S_{\rm in}(\xi) =\gamma.
\end{equation}
For example  for  equation (\ref{eq:2}), 
\begin{equation}
  \label{eq:16b}
g(\rho)=\frac{1}{1+K} 
\left ( 2\rho+\frac{K-1}{1+K}\log [K - (1+K)\rho]\right ),
\end{equation}
and  $S_{\rm  in}(\tilde x) \sim \tanh (\tilde x/(2w) +\xi)$, for all $K$.
The two scales mentioned in equations (\ref{eq:4}) and (\ref{eq:4w}) appear
in the inner solution.   There is an $N$-dependence of $\xi\sim
N^{-\nu_{-}}$ with an exponent $\nu_{-}=1$ to make $\xi$ act also as the
finite size scaling variable for the transition.  

The condition for saturation of $S_{\rm in}(\tilde x)$ is 
\begin{equation}
F(\rho)=0 \ {\rm for}\ \rho=\rho_s >\rho_{\rm o}.
\end{equation}
For $\gamma>\rho_{\rm in}(\tilde{x}\to\infty)$ the boundary condition
cannot be satisfied by $\rho_{\rm in}$ leading to shock formation.
Therefore the phase boundary is given by $\gamma=\rho_s(\rho_{\rm
  o}(\alpha_s))$ with shocks appearing for $\alpha >\alpha_s(\gamma)$.
We note here that since $F(\rho)$ has a vanishing derivative at
$\rho=\rho_c$, by Rolle's theorem of calculus, $\rho_{\rm o}\leq
\rho_c\leq\rho_s$.  In the phase with shock, the center of the shock ,
can always be chosen such that $\rho_{\rm in}(0)=\rho_c$ at $x_d=x_s$

Assuming simple zeros at $\rho=\rho_{o}$ and $\rho=\rho_{s}$, we
write
\begin{equation}
F(\rho)=-(\rho-\rho_{\rm o})(\rho-\rho_s)\ \phi(\rho),
\label{eq:frho}
\end{equation}
which defines $\phi(\rho)$.  For the the case of equation (\ref{eq:2}),
$\phi(\rho)=1$.   
The large  $\tilde x$ behavior is then given by
\begin{equation}
  \label{eq:6}
  \frac{d\rho}{d\tilde x}\approx -  \frac{\rho-\rho_s}{w(\alpha)},
\ \mathrm{where\ }
w(\alpha)=(\rho_s-\rho_{\rm o})^{-1} 
\ \frac{f_2(\rho_s)}{\phi(\rho_s)},
\end{equation}
with  $w$ depending only on $\alpha$ and not on $\gamma$.
equation (\ref{eq:6}) shows $w(\alpha)$ as the characteristic length
scale for approach to saturation or the bulk density, as defined
earlier.  A similar equation describes the approach to $\rho_{\rm o}$ with a
scale $w_0\propto w$.
  This allows a practical definition of $w$ as
\begin{equation}
  \label{eq:9}
  w^{-1}=-\left. {d \ln (\rho(x)-\rho_{\rm out}(x))}/{d \tilde x}\right|_{\tilde
    x\to -\infty},
\end{equation}
noting that $\rho_{\rm out}(x)$ is the bulk density and the deviation
is only in the boundary layer region.

The behavior of $\xi$ near the phase boundary, for a fixed $\gamma$
can be determined from the condition at ${\tilde{x}}=0$, equation
(\ref{eq:3}).  For $\gamma$ greater than but close to $\gamma_c$, equation
(\ref{eq:14}) together with equation (\ref{eq:frho}), yield
\begin{equation}
\xi\sim \ln |\gamma-\rho_{\rm s}|\sim \ln|\Delta\alpha|,\quad {\rm or}
\quad \zeta_{-}=0 (\log),
\end{equation}
from equation (\ref{eq:4}). This scale $\xi$ depends on both the boundary
conditions.  To be noted that 
the shockening transition from the low-density 
side determines the phase boundary.

For a given $\gamma$, $w$ may be made to diverge by tuning $\alpha$.
This locates the critical point on the phase boundary  at
$(\alpha_c,\gamma_c)$ such that
\begin{equation}
  \label{eq:11}
g(\gamma_c)-g(\alpha_c)=\Omega,\ {\rm with}\ \rho_{\rm o}=\rho_s=\rho_c=\gamma_c.
\end{equation}

If the boundary condition at $x=1$ is held fixed at $\gamma=\gamma_c$,
then for $\alpha\to \alpha_c$, both $\rho_{\rm o}$ and $\rho_s$
approach $\rho_c$.  For $\rho_{\rm o}=\rho_c -\delta\rho$ and
$\alpha=\alpha_c -\delta \alpha$ one may expand equation (\ref{eq:16}) in
$\delta\rho$, $\delta \alpha$ with $x=1$.  Now, by definition,
$g^{\prime}(\rho)= f_1(\rho)/f_0(\rho)$, so that
\begin{equation}
  \label{eq:8}
%g^{\prime}(\rho)= f_1(\rho)/f_0(\rho),
%\ {\rm so\  that}\  
g^{\prime}(\rho_c)=0,\ {\rm if}\  \rho_c\neq \rho_L.  
\end{equation}
We therefore have $w\sim|\rho_s-\rho_{\rm o}|^{-1} \sim
|\Delta\alpha|^{-\zeta_c},$ where
\begin{equation}
\label{eq:5}
\zeta_c=1/2, \  {\rm if}\ \rho_c\neq \rho_L, \quad {\rm but}\quad 
 \zeta_c            = 1, \   {\rm if}\ \rho_c =\rho_L.
\end{equation}
The density for the critical point also shows a singular variation
near the end point, namely, $\rho_c-\rho(x) \sim \sqrt{1-x}$ for $x
\to 1-$, if $\rho_c\neq\rho_L$.  This follows from equation (\ref{eq:14})
as a consequence of the simple zero of $f_1(\rho)$.  These results can
be verified for the special case of equation (\ref{eq:2}) but are seen here
to be of more general validity.

The shape of the phase boundary can be determined in a similar way.
For $\gamma=\gamma_c+\Delta\gamma$, equation \ref{eq:16}, with $x=1$, on
expansion gives
$$-g^{\prime}(\rho_c) \Delta\gamma +
g^{\prime\prime}(\rho_c)(\Delta\gamma)^2/2+ ...
=g^{\prime}(\alpha_c)\Delta\alpha ... .$$
The phase boundary therefore
takes the form $ \Delta\gamma\sim |\Delta\alpha|^{\chi_{-}}$ with
\begin{eqnarray}
  \label{eq:26}
  \chi_{-}&=&1, \quad {\rm if}\ g^{\prime}(\rho_c)\neq0 \quad 
({\rm e.g.} \ K=1)\\
  &=&1/2, \quad {\rm if}\ g^{\prime}(\rho_c)=0, \quad 
({\rm e.g.}\ K\neq1)
\end{eqnarray}
where $K$ refers to equation (\ref{eq:2}).  The analogous exponent
$\chi_{+}$ for $\gamma<\gamma_c$ is discussed later.
 
Though $w$ measures the crossover length, right on the phase boundary,
it is related to the height of the shock that forms beyond the phase
boundary, namely, $h=\rho_s-\rho_o\sim w^{-1}$.  Mean field results,
based on power series expansion, seem to require $\chi_{-}\leq 1$ so
that the phase boundary is not tangential to the $\alpha$-axis.  This
implies that $\Delta\gamma$ can be taken as a measure of distance $r$
from the critical point measured along the phase boundary, i.e.,
$r\sim\Delta\gamma$.  Along this curve, for $\gamma\geq \gamma_c$ or
$r\leq 0$,
\begin{equation}
  \label{eq:1}
h\sim |r|^{\beta} \ {\rm with}\  \beta=1.  
\end{equation}
For $\gamma<\gamma_c$, the length $w$ remains infinite, i.e. the
shock, if formed, is of zero height.  In fact, the boundary layer that
forms does not shocken (path 3 in Fig. \ref{fig:sch}).  In this
situation, there are two possibilities, either the shock height grows
continuously as the phase boundary is crossed or the shock formation
is suppressed.  Though the former is the generic scenario, the latter
situation can occur under special conditions like e.g. the special
symmetry when $\rho_c=\rho_L$.

We now show that the same properties of the $f$-functions also
determine the behavior in the deconfined region.  The shock state,
just after deconfinement, is described by the thickness of the
deconfined layer measured by the distance of the shock from the
boundary, $\Lambda$, and the height, $h$ of the shock.  If the
deconfinement takes place at $x=1$ and the shock position is $x_s$,
then $\Lambda=1-x_s$.  In addition the shock layer will have a width
which we do not discuss here.  The behavior of $h$ along the phase
boundary is also of importance, especially as one approaches the
critical point.

To discuss shock, we need the outer solution of equation  (\ref{eq:14}) 
satisfying $\rho(1)=\gamma$ as
\begin{equation}
\Omega (1-x)=g(\gamma)-g(\rho_{\rm out})  \quad ({\rm right \ solution})
\label{eq:16a}
\end{equation}
Moreover, the shock close to the critical point is asymptotically
symmetric with a scale $w=(\rho_s-\rho_{\rm o})^{-1}
f_2(\rho_c)/\phi(\rho_c)$ as seen from equation (\ref{eq:6}).  For a
symmetric shock, centered at $\rho=\rho_c$, its position $x_s$ and
height $h$ satisfy equations (\ref{eq:16}) and (\ref{eq:16a}) with
$\rho_{\rm out}=\rho_c \mp h/2$ respectively.  By expanding equations
(\ref{eq:16}), and (\ref{eq:16a}) in $h$ and $\Delta\alpha=\alpha
-\alpha_c $ keeping $\gamma=\gamma_c$, we get (prime denoting
derivatives) (path 2' in Fig. \ref{fig:sch})
\begin{equation}
\label{eq:18}
 g^{\prime}(\rho_c) h + \frac{1}{24}g^{\prime\prime\prime}(\rho_c) h^3 + ... 
=  -g^{\prime}(\alpha_c) \Delta\alpha + ... .
\end{equation}
A similar analysis for $x_s$ can also be done.  In the general case,
$\rho_c\neq\rho_L$, if the third derivative of $g$ is non-vanishing,
equation (\ref{eq:18}) gives
%\begin{subequations}
\begin{equation}
  \label{eq:7}
 h \sim  | \Delta\alpha|^{\beta^{\prime}},\ {\rm with}
\ \beta^{\prime}=1/3,\ {\rm and} \ \Lambda=1-x_s\sim | \Delta\alpha|^{\zeta},
\ {\rm with}\ \zeta=2\beta^{\prime}.  
\end{equation}
%\end{subequations}
equation  (\ref{eq:7})  defines the bulk exponents
$\beta^{\prime}$ and $\zeta$ for $h$ and $\Lambda$ respectively.
These exponents have been found for equation (\ref{eq:2}) with $K\neq 1$
\cite{frey}, but is shown here to be more general. There are other
possibilities also.  E.g. in case all the derivatives of $g(\rho)$
vanish, no shock can exist\cite{evans}, as, e.g., for $K=1$ in equation
(\ref{eq:2}).  Such special cases will be discussed elsewhere.  The
exponents obtained in equation (\ref{eq:7}) remain the same for all
$\gamma\leq\gamma_c$ (path 3' in Fig. \ref{fig:sch}) because in this
regime the effective boundary condition is $\gamma=\gamma_c$, thanks
to the formation of a nonshockening boundary layer at $x=1$.  A
consequence of this is that $\chi_{+}=0$ for the shape of the phase
boundary for $\gamma<\gamma_c$.  For $\gamma>\gamma_c$ (path 1' in
Fig.  \ref{fig:sch}), $h$ remains O(1) on the phase boundary so that
$\beta^{\prime}=0$ and $\zeta=1$.  It is tempting to suggest a scaling
relation $\beta^{\prime}+\zeta=1$ throughout.

As a further example of the predictive power of this approach, we
consider ASEP of interacting particles \cite{kls,schmitt} that
destroys the particle-hole symmetry.  Our aim is to show that the same
physical picture remains valid quantitatively in this case also.  In
the interior, particles at site $i$ move to site $i+1$, provided it is
empty, with a rate that depends on the state of sites $i-1$ and $i+2$
as 0100$\rightarrow$ 0010 with a rate $1+\delta$ and 1101$\rightarrow$
1011 with a rate $1-\delta$, while all other rates remaining the same
as for equation  (\ref{eq:12}).  For equal attachment and detachment rates
($K=1$), the continuum mean-field equation describing the shape of the
density profile is of the type equation (\ref{eq:25}) with $f_2(\rho)= 1+
\delta(1-2\rho)$, $f_1(\rho)=1-2\rho+\delta[1-6 \rho(1-\rho)]$ while
$f_0(\rho)$ remains same as in equation (\ref{eq:2}) with K=1, i.e.  with
$\rho_L=1/2$.  Numerical (MATLAB) solution of the differential
equation for relevant set of parameters actually shows that the shock
is centered at $\rho_c= (1+3\delta-\sqrt{1+3\delta^2})/(6\delta)$, the
zero of $f_1(\rho)$.  In particular, $\rho_c<1/2$.  For the shockening
transition, keeping $\gamma$ fixed, the relevant zero, $\rho_s$, of
${\hat f}_1(\rho_{\rm o})-{\hat f}_1(\rho)$ is ($\sigma=2\rho_{\rm o}
-1$)
\begin{equation}
  \label{eq:10}
4\delta \rho_s= 
{1+2\delta-\delta \sigma -(1-2\delta^2+2\delta\sigma
  +3\delta^2\sigma^2)^{1/2}}.
\end{equation}
The phase boundary between the low-density and the shock phase is
determined by the condition $\gamma=\rho_s(\rho_{\rm o},\delta)$. The
height of the shock on the phase boundary is $\gamma-\rho_{\rm o}$
which vanishes at the critical point $\gamma_c=\rho_c$.  It follows
from these that the shock near the critical point is asymptotically
symmetric.  So the critical feature, especially of the shockening
transition of this interacting system is similar to the general case
considered above (except for $\delta=0$).  This can also be explicitly
verified from detail solutions.

In summary, we have shown the existence of universal behavior
associated with shockening transition, a precursor to the shock
formation.  The dynamics is described by a set of $f$ functions as
defined by equation (\ref{eq:25}).  The zeros $\rho_L$ and $\rho_c$ of
$f_0(\rho)$ and $f_1(\rho)$, respectively, are the two important
densities for the steady state.  The shock, when formed, is centered
around $\rho=\rho_c$ which is a point of special symmetry. The generic
situation corresponds to $\rho_c\neq\rho_L$.  The thickening of the
layer that leads to the shock (called the ``shockening layer'') is
described by two diverging scales, $\xi$ and $w$ whose critical
behaviors are determined by the nature of $f_i(\rho)$ around
$\rho=\rho_c$.  Furthermore, our analysis and results suggest the
possibility of more complex dynamics involving different symmetries,
though such model systems are not known till now.

SM acknowledges  financial support from IIT, Kanpur and thanks Bhaskar
Dasgupta for fruitful suggestions. 

%\vspace{ -0.5cm}

\end{document}